# Molecular Simulation of a Zn–Triazamacrocyle Metal–Organic Frameworks Family with Extraframework Anions

*Marta De Toni,[a,b] François-Xavier Coudert,\*[a] Selvarengan Paranthaman,[a]*

*Pluton Pullumbi,[b] Anne Boutin,[c] Alain H. Fuchs[a]*

[a] *Chimie ParisTech & CNRS, UMR 7575, 11 rue Pierre et Marie Curie, 75005 Paris, France.*

[b] *Air Liquide, Centre de recherche Claude Delorme, 78354 Jouy-en-Josas, France.*

[c] *Chemistry Department, École Normale Supérieure, CNRS, UPMC, Paris, France*

E-mail: fx.coudert@chimie-paristech.fr

ABSTRACT. We report an investigation by means of adsorption experiments and molecular simulation of the behavior of a recently synthesized cationic metal–organic framework. We used a combination of quantum chemistry calculations and classical, forcefield-based Grand Canonical Monte Carlo simulations to shed light into the localization of extra-framework halogenide anions in the material. We also studied the adsorption of small gas molecules into the pores of the material using molecular simulation, and investigated the coadsorption of binary gas mixtures.

Nanoporous metal–organic frameworks (MOFs) are topical materials displaying a large range of crystal structures and host–guest properties, due to a combination of tunable porosity, by choice of metal centers and linker length, and functionalisation of the internal surface of the material. Among the proposed applications of MOFs, adsorptive separation of strategic gases ($H_2$, $CO_2$, $CH_4$, …) is of particular importance and has gained a lot of attention in recent years. In particular, a large number of MOFs have been benchmarked for selective adsorption of $CO_2$ in $CO_2$/$CH_4$ mixtures.[1,2,3] The quest for the design of better adsorbents with a fine tuning of pore size, pore shape and chemical functionalisation has lead to the development of entire families of metal–organic frameworks, based on common metal centers and linkers sharing certain characteristics. Examples of such families include the IRMOFs,[4] based on $Zn_4O$ clusters and dicarboxylate linkers, the zeolitic imidazolate frameworks (or ZIFs),[5] as well as pillared materials using both carboxylate- and pyridine-based linkers. Besides these examples, the recent family of zeolite-like metal–organic frameworks (ZMOFs)[6] contains anionic MOFs which, by way of charge compensation, feature exchangeable extra-framework cations. These cations, which are typically accessible by guest molecules diffusing in the pores of the host matrix and may be substituted after the material has been synthesized by simple ion exchange, make the ZMOFs promising materials for hydrogen binding.

In this article, we focus on a new type of cationic metal–organic framework recently synthesized by Ortiz et al.[7] based on a triazamacrocycle and carboxylate groups binding $Zn^{2+}$ cations, which features $Cl^-$ anions in its nanopores to compensate the net positive charge of the bare framework. In particular, we used molecular simulation of this new family of materials to investigate the localization of $Cl^-$ and other extra-framework anions, which has not yet been experimentally determined. One of the motivations for this study of anion localization is that, in the well-studied family of zeolites, it was shown that the distribution of extra-framework cations is required to understand gas adsorption and catalytic activities. As a consequence, the characterization of the distribution of anions inside the pores of positively-charges MOFs is of importance for their practical applications. In addition to this, we report both theoretical simulations and experimental measurements of the adsorption of $CO_2$ in this new MOF, and study $CO_2/CH_4$ separation by means of a molecular simulation coadsorption study.

## 1. System: the Zn-CBTACN metal–organic framework

The MOF under study is built from $Zn^{2+}$ cations as metal centers, linked with triazamacrocycle substituted with carboxylic groups as a linker. This organic linker is 1,4,7-tris(4-carboxybenzyl)-1,4,7-triazacyclononane (henceforth called CBTACN), which is an N-substituted triazacyclononane, represented in Fig. S1. The chemical formula for the dehydrated crystalline compound is $[Zn_2(CBTACN)]_{16}$.[7] It has a highly symmetric cubic structure, with space group $I\bar{4}3d$ and a unit cell parameter of $a = 25.86$ Å. A unit cell of Zn-CBTACN is represented in Fig. 1. It presents one-dimensional channels of ~8 Å diameter, colinear with the crystallographic axes and perpendicular to each other. These channels are interconnected two by two, with windows of 5.5 Å in diameter. Its Langmuir surface area was measured at 1350 m²/g, and the BET surface is 1199 m²/g.

The organic linker is bound to the $Zn^{2+}$ dimers in an asymmetric fashion: one of the zinc ions is chelated by the three nitrogen atoms of the triazamacrocycle, and bonded to one oxygen atoms of each of three carboxylate groups, while the second zinc atom of the cluster is bonded to the remaining three oxygen atoms of the same carboxylate groups (see Fig. 2). Thus, the first Zn cation has an octahedral environment, while the second is at the center of a tetrahedron; the apex of this tetrahedron, which is not occupied in Fig. 2, is the location of a coordinated water molecule in the as-synthesized material. Upon activation of the framework, this apical water molecule can be removed and the dry material is obtained. It is on this ideal, water-free, crystalline structure that molecular simulations were performed.

## 2. Molecular simulation forcefield

### 2.1. Adsorbate forcefields

All Monte Carlo calculations in this work were performed using atomistic models of the MOF framework and adsorbate molecules. The positively charged Zn-CBTACN framework was considered rigid, while the charge-balancing extra-framework anions were fully mobile. The adsorbates were modeled as rigid molecules, using standard forcefields that have demonstrated good quality for reproducing phase equilibrium and structural fluid properties. Carbon monoxide (CO) was described by a model with two Lennard-Jones centers, on the atoms themselves, and three point charges (a positive partial charge of $+1.7\ e$ in the middle of the bond and two negative charges : one of $-0.8\ e$ on the C atom and one of $-0.9\ e$ on the O atom).[8] The apolar $CH_4$ molecule was described by a single force center.[9] Water molecules were described by the TIP4P-Ew model,[10] a modified TIP4P potential for use with the Ewald summation technique, which was shown to present good thermodynamic properties for bulk water.[11]

The $CO_2$ molecules were initially described by the TraPPE model, which presents three Lennard-Jones force centers and three point charges (a positive partial charge of $+0.7\ e$ on the C atom and a negative one of $-0.35\ e$ on each O atom).[12] As detailed in the Results section, these initial simulations using a standard forcefield failed to describe adequately the $Cl^-$–$CO_2$ interactions, and a custom nonelectrostatic forcefield was designed (see Section "$CO_2$ adsorption: molecular simulation").

**2.2. Determination of atomic partial charges by quantum chemistry calculations**

We now turn to the description of the forcefield describing the metal–organic framework's interactions with the adsorbed species. Because the Zn-CBTACN material studied here is built from quite unusual triazamacrocycle-based linkers, with coordination modes that are, to our knowledge, not found in materials previously reported and studied in the literature, we could not adjust or modify existing forcefields from related materials. We thus needed to construct a forcefield describing Zn-CBTACN from scratch. Following the most common practice in the field,[13] we combined a standard forcefield for the description of repulsion and dispersion interactions with partial point charges on the MOF's atoms to reproduce its Coulombic interactions.

The repulsion and dispersion energies within the material were modeled by a Lennard-Jones potential and its parameters come from the widely used DREIDING force field (whose relevant parameters are reported in Table S1).[14] The interactions between the Zn-CBTACN material and the molecules adsorbed described above where determined by the Lorentz-Berthelot mixing rules. In addition these repulsion–dispersion interactions, host–guest interactions in the system need to include electrostatic terms, which were described as a Coulombic interactions between sets of point charges borne by the framework, the mobile anions and the adsorbed species. Atomic partial charges for the halogenide anions were set to $-e$, following the common practice of modeling cations in zeolites, which can be reasonably described with monovalent cations bearing an net charge of $+e$ (ref. 15 and citations therein). Partial charges of the guest gas molecules were taken from their respective models, as described in the previous section.

The partial charges of the metal–organic framework were determined from quantum chemistry calculations on a representative cluster extracted from the crystalline structure of Zn-CBTACN. Various strategies have been used in the community in recent years to obtain partial atomic charges for porous solids, including charge-equilibration methods (long used for zeolites, and more recently introduced for MOFs[16]) as well as quantum chemistry calculations. The former take into account the whole periodic framework of the material, while the computational cost of quantum-based methods forbids their use for full-cell periodic calculations in materials featuring large unit cells. Recent publications focusing on cluster calculations for MOF building blocks[17] have validated the use of clusters of increasing size by comparing the results to periodic calculations. The cluster chosen in this work contains 122 atoms ($Zn_2C_{54}H_{51}O_{12}N_3$); it is represented in Figure 2. It is hydrogen terminated for each carbon atom which would normally be linked to a further triazamacrocycle, while the terminating carboxylate groups (which would bind to a zinc dimer unit) were left unprotonated to more accurately describe the electronic distribution of the crystalline, periodic structure. The cluster was first optimized at the HF/6-31+G(d) level, after which we used density functional theory calculations with the same 6-31+G(d) basis set and the PBE0 exchange-correlation functional.[18] Electrostatic charges were obtained using the ChelpG method,[19] which has been widely and successfully used to obtain atomic partial charges in a wide variety of metal–organic frameworks.[20,21,22] The calculations were performed using the Gaussian 03 package,[23] and the charges obtained were averaged for groups of symmetry-related atoms. The resulting partial charges, which were then used throughout the Monte Carlo simulations, are reported in Table 1.

### 3. Localization of extraframework anions

### 3.1. Chloride anion

The Zn-CBTACN MOF synthesized by Ortiz et al. has chemical formula $[Zn_2(CBTACN)Cl]_{16} \cdot (H_2O)_{16}$. The Zn-CBTACN framework itself is positively charged. As a consequence, there exist charge-balancing counter-anions in the material which, in the case of the material under study, are chloride ions $Cl^-$. These anions, which could not be localized from the X-ray diffraction data recorded by Ortiz et al., are thus expected to be relatively disordered and not ionocovalently bound to the framework. We call these $Cl^-$ *extra-framework anions*. They are required as part of the material for electroneutrality reasons, yet are not part of the crystalline structure itself, and they are expected to play an important role in the physicochemical properties of Zn-CBTACN and other related materials. We hypothesize that the accessibility of anions to guest molecules will be strongly related with adsorption properties and catalytic activity of the materials. As a consequence, it is of importance to study the location (or distribution) of anions inside the Zn-CBTACN to better understand the properties of this MOF. No experimental data is yet available on this issue of anions localization, although elemental analysis has proved that these anions are indeed present in the activated material. We show here that molecular simulation allows to get insight into their distribution.

We studied the distribution of three different anions in the framework of Zn-CBTACN: the chloride anion, which is the anion experimentally present in the material synthesized and reported by Ortiz et al., and the bromide and fluoride anions, which we studied in order to analyze the influence of anion size and polarizability on the ion distribution. The unit cell of bare Zn-CBTACN has framework charge +16, so that if the counter-anions are monovalent, as are halogenides, there are 16 extraframework anions per unit cell. This is the same number as that of accessible $Zn^{2+}$ cations in the structure, or coordinatively unsaturated sites (CUS). The undercoordinated $Zn^{2+}$ cation is represented in dark green on the cluster of Figure 2; in the view represented there, it is fully accessible from the top.

The localization of all three halogenide ions were studied by means of molecular simulation in the ($N$, $V$, $T$) ensemble, including local translations and random displacement moves. The anion distributions were calculated for three different temperatures, 77 K, 300 K and 1000 K, in the guest-free Zn-CBTACN framework. For each anion and temperature, long equilibration runs of 400 million MC steps were performed, leading to a well-equilibrated distribution of the anions could be found. The convergence was systematically checked by comparing distributions obtained from simulations with widely different initial positions of the anions.

The probability densities for all three anions studied were found to be discrete, i.e. these distributions are composed of individual probability clouds, separated from one another by regions of zero density, were no anion was observed during the full length of a simulation. Anionic sites are related to one another by the symmetry operators of the crystal group of Zn-CBTACN, yielding for each anion and each temperature a single anionic site, with a well-defined position and spread. The chloride anions, which are the ones found in the experimentally synthesized material, are located in the principal channels of the unit cell, which are aligned with the crystallographic axes $a$, $b$, and $c$. They are localized in-between two channel intersections (see Figure 3), and not inside the intersections themselves, even though their kinetic radius of 1.9 Å is close to the geometric size of the opening connecting neighboring channels. We interpret this as a way for these anions to maximize their dispersive interactions with the surrounding atoms of the material, by being inside the channel rather than at an opening.

In addition to the issue of site localizing, the distribution of Cl- anions among the identified symmetry-equivalent sites is a nontrivial issue, as it can be for extraframework cations in zeolites.[24] Two chloride ions belonging to the same channel are separated by a distance of at least 6 Å (and at most 12 Å), while anions across the window connecting two neighboring perpendicular channels would feature a much smaller distance of 4–5 Å. As a consequence, the Cl– anions are statistically distributed in all the channels in a manner to minimize their strong electrostatic repulsion as, again, was found to be the case of extraframework cations in zeolitic materials.[24,25] In the structure of Zn-CBTACN, there are 12 channels in each unit cell, with 4 channels parallel to each crystallographic axis (see Figure 1). We found that, at all temperatures studied (from 77 K to the unrealistic 1000 K), the distribution of anions among the various channels was quite even, with the most probable distribution being (6,5,5), i.e. 5 anions in each system of channels, except for one orientation which has 6 anions. The occurrence of other observed distributions, which are (6,6,4) and (7,5,4), are of much lower probability. They strongly depend on temperature, with their cumulative percentage growing from 5% at 77 K, to 15% at 300 K (and 20% at 1000 K).

### 3.2. Localization of other extra-framework anions

Having discussed the localization of the $Cl^-$ ion in the porous space of framework, we now turn our attention to the influence of anion size on localization. Thus, we studied how two other halogenides, $F^-$ and $Br^-$, would be distributed inside the pores. The latter, $Br^-$, occupies sites very close to those of $Cl^-$, but slightly shifted away from the walls of the framework, and with a smaller spread. This can be explained by the larger radius of the bromide anion (which has a kinetic radius of 2.3 Å), and by the fact that $Br^-$ could not, in any case, fit in the channel intersections, which have a ~4 Å opening.

The smaller $F^-$ ion, however, has a very different behavior from the other two. We found that the fluoride anions are localized inside more condensed sites, and are remarkably less mobile (in terms of the spread of probability density) than $Cl^-$ and $Br^-$. While the spread of the $Cl^-$ site is around 3 Å in diameter, the $F^-$ anions are located in a very narrow region near the undercoordinated $Zn^{2+}$ ions. In fact, the fluoride anions are well-ordered, with one $F^-$ per exposed $Zn^{2+}$, in a very well-defined geometry, described in Figure 4. In this geometry, the fluoride anion is at a distance of 2.7 (± 0.1) Å of the zinc (II) ion, and forms an angle of ~ 30° with the Zn–Zn axis. This position allows it to be in proximity to one aromatic ring of the framework, maximizing dispersive interactions while maintaining a strong electrostatic interaction with the undercoordinated metal center. These sites, while very favorable for an anion to "lock in", are inaccessible to the $Br^-$ and $Cl^-$ ions due to their larger ionic radius.

As a consequence of these differences we can hypothesize that, if the fluoride version of Zn-CBTACN were to be synthesized, it is likely that the F$^-$ anions could be localized by X-ray diffraction techniques, unlike chloride anions whose distribution has not been determined. We will also, in the following, show how this quite different anion distribution impacts physicochemical properties of the material.

## 4. Gas adsorption experiments and simulations

### 4.1. Adsorption experiments

The Cl-Zn-CBTACN sample has been activated before gas-sorption analysis. The activation protocol for the MOF consisted in a combination of controlled temperature increase (7 hours from room temperature to 423 K and 5 hours at 453 K) and secondary vacuum (10$^{-5}$ mbar). Low-pressure gas-sorption experiments (up to 1 bar) were performed on a Micromeritics ASAP 2405 volumetric instrument. High-pressure adsorption isotherms for $CO_2$ and CO (up to 30 bar) were performed on a HPA-400 volumetric instrument (VTI, Hialeah, FL). The compressibility factors of high-pressure gases were determined by using the REFPROP program[26] and the NIST Standard Reference Data Base 23.

### 4.2. $CO_2$ adsorption: experimental results

The initial report of the synthesis of Cl-Zn-CBTACN, by Ortiz et al., included a characterization by means of room temperature adsorption of $CO_2$, CO, $N_2$ and $O_2$ at low pressure (up to 1 atm). We report in Figure 5 the adsorption isotherms of $CO_2$ taken successively at 274 K, 303 K, 333 K, measured in this order on the same sample. The figure also includes a second isotherm at 274 K, taken after the 333 K one, and demonstrating the good stability of the material and the good reproducibility of the activation procedure followed (see the "Methods" section). It can be seen this high-pressure data is in remarkably good agreement with the low-pressure isotherm reported in ref. 7 (plotted in black on Figure 5). In particular, the Henry constant of both isotherms (i.e. the slope of the isotherm at P → 0) have the same value of 0.7 kPa$^{-1}$. This value, which is higher than those of $CH_4$ (0.17 kPa$^{-1}$) and CO (0.03 kPa$^{-1}$) at room temperature, hints at a strong potential for separation of $CO_2$ from these gases, which could be used in removal from flue gas ($CO_2/CO$) or sour natural gas wells ($CO_2/CH_4$). To confirm these properties, we have thus performed molecular simulation of pure component and mixture adsorption in Zn-CBTACN.

**4.3. Grand Canonical Monte Carlo simulations**

Adsorption properties of various guest molecules ($H_2O$, $CO_2$, CO and $CH_4$) in Zn-CBTACN were studied using atomistic Monte Carlo simulations. Adsorption isotherms were computed by a series of Grand Canonical Monte Carlo simulations, with each point of the isotherm obtained by a single GCMC calculation at fixed chemical potential. The chemical potential was then related to the adsorbate vapour pressure by using an ideal gas law for most adsorbent, which is valid in the range of pressure used for all adsorbates but $CO_2$. In this case, we used the experimental fugacity–pressure relationship as obtained from the NIST fluid database.[26] Furthermore, the absolute adsorbed quantities obtained directly from the GCMC simulations were converted into excess adsorption,[27,28] which can be directly compared with experimental data.

For mixture coadsorption, the external partial pressure of each gas was used to determine its chemical potential in the same manner. The adsorption selectivity for a A:B mixture at a given pressure was calculated as $\rho_{A:B} = (y_A/y_B)/(x_A/x_B)$, where $x_i$ is the molar fraction of component $i$ in the external fluid, and $y_i$ is the molar fraction in the adsorbed phase. GCMC simulations were typically run for 10 million steps on a single unit cell of the host material, which has 1136 atoms. Long-range electrostatic interactions were taken into account using the Ewald summation technique. In order to improve the efficiency of the calculations, the electrostatic and repulsion–dispersion interaction energies between the rigid MOF framework and both extra-framework cations and adsorbates were precomputed on a grid (with a grid mesh of 0.15 Å) and stored for use during the simulation. Preinsertion, orientation and jump biased moves were used to accelerate the convergence of the Monte Carlo simulations. Moreover, the heats of adsorption of the various guests were also determined in the limit of zero loading. They were calculated by Monte Carlo simulations in the Grand Canonical ensemble at low coverage ($<N_{ads}> \sim 1$) from the fluctuations of the intermolecular energy in the adsorbed phase and of the number of adsorbed molecules.[29,30]

**4.4. $CO_2$ adsorption: simulation results**

We report in Figure 6 the adsorption isotherm of $CO_2$ computed using molecular simulation at 274 K, 300 K and 334 K, using the TraPPE model for $CO_2$. While the trend of the adsorption isotherms with temperature is coherent, the shape of each individual simulated isotherm does not match with that of the experimental one. The saturation uptakes calculated from molecular simulation are within 5 to 10% of the experimental values. However, a significant difference in the very low pressure region of the isotherm is clear, and leads to theoretical Henry constants much higher than those measured experimentally. Many factors could be explain these differences, and we first investigated the possible effect of traces of water in the material on the adsorption of $CO_2$ by running GCMC simulations run in presence of 16 water molecules per unit cell, i.e. one per crystallographic water site in the as-synthesized material (which was deemed an upper bound on the possible hydration of the activated sample). The resulting isotherm at 300 K is shown in Fig. S3. It has a lower Henry constant due to water molecules binding themselves to the extraframework $Cl^-$ anions (3.5 kPa$^{-1}$ instead of 4.5 kPa$^{-1}$).[31] The scale of this change is, however, too small to account for the difference with the experimental results, and this explanation is not sufficient. Moreover, the addition of water molecules reduces the $CO_2$ saturation uptake, leading at high hydration to values incoherent with the experimental isotherms.

As a second hypothesis on the reason behind this discrepancy, we tested the validity of the molecular forcefields used. Decomposition of the adsorption enthalpy for $CO_2$ revealed that the excessively high Henry constant found using this first forcefield was mainly due to the electrostatic $Cl^-$–$CO_2$ interactions. We attribute this to the fact that the charges in the TraPPE $CO_2$ molecule, which are adequate to reproduce $CO_2$ in the condensed liquid and supercritical states, appear unsuitable to describe the interaction of a single isolated molecule with a chloride anion. Indeed, in situations of severe confinement or in the presence of few adsorbed molecules, the dipole of confined adsorbates can vary quite drastically.[32,33,34] Molecular simulation using nonpolarizable forcefields do not cope well with this situation, leading to the need of "turning down" the dipole moment of molecules from its values optimized to describe the bulk phase.[30] This approach has been used in the existing literature for highly polar molecules, such as $H_2O$[30] and $CO_2$.[35] In the case of $CO_2$ adsorption in Cl-Zn-CBTACN, we ran GCMC simulations with various values of the $CO_2$ quadrupole moment. It turned out that it is necessary to turn off MOF–$CO_2$ and Cl⁻–$CO_2$ electrostatic interactions completely in order to obtain the correct adsorption enthalpy for $CO_2$, and thus the correct Henry constant. However, doing so worsens the value of saturation uptake because the density of the condensed adsorbed phase is not described appropriately by a purely Lennard-Jones $CO_2$ potential. We have thus re-adjusted the two Lennard-Jones parameters at the same time, to reproduce the adsorbed density in the pores at saturation. We also checked that the resulting forcefield yields a reasonable description of the bulk condensed phase. The resulting parameters are given in Table 2. It is worth mentioning that equivalent results can be obtained by turning off the partial charges of the host framework for interaction for $CO_2$ molecules.

Thus, we run a second series of molecular simulation using this nonelectrostatic $CO_2$ molecule. This model yields an excellent agreement between simulated and experimental isotherms at all temperature, for both the low-pressure slopes, the saturation uptakes and the overall shapes of the isotherms (Fig. 7). The heats of adsorption, calculated from low-pressure GCMC simulations, is of $24 \pm 2$ kJ.mol$^{-1}$ in the temperature range explored. It is in good agreement with the experimental values, as calculated by the variation of the Henry constant with the temperature, of $21.5 \pm 0.5$ kJ.mol$^{-1}$. However, the readjustement of the forcefield necessary to obtain these results highlights a general problem in the simulation of adsorption of polar molecules in metal–organic frameworks: atomic partial charges derived from quantum chemistry calculations can not always be mixed freely with adsorbate forcefields optimized for bulk condensed phase. This issue requires a more rigorous investigation in future work.

### 4.5. Gas separation in materials of the X-Zn-CBTACN family

In order to evaluate the potential of Cl-Zn-CBTACN for adsorptive gas separation, Ortiz et al. calculated in ref. 7 the low-pressure selectivity of CO, $O_2$ and $N_2$ over $CO_2$ from the ratio of Henry constants from pure component isotherms. We used molecular simulation to make predictions and help provide some broader perspective into the issue of gas separation in this family of halogenide-Zn-CBTACN. Thus with the intent to show concretely the interest of this new MOF by its selectivity, the adsorption of $CO_2/CH_4$ gas mixtures with different molar fraction were studied. The coadsorption isotherms are shown in Figure 8 for the 50:50 mixture. As expected, the relatively strong interactions of $CO_2$ with the framework and the anions mean that it adsorbs in larger quantities than methane. The selectivities (displayed as a function of pressure in Fig. 9) grow from 3.2 at low pressure, which is equal to the ration of Henry constants, to a limiting value of 10 at $P(CO_2) \sim P_{sat}$, which is 71 bar at 303 K. In addition, it is seen in Fig. 9 that the selectivity towards the $CO_2/CH_4$ mixture is roughly independent from the molar fraction of the mixture. This property, while always true asymptotically at low pressure, is not so common in microporous materials which tend to have pressure-dependent selectivities.

We also studied the influence of the nature of the extraframework anion on the separation properties of the X-Zn-CBTACN for X = $F^-$, $Cl^-$ and $Br^-$. The coadsorption isotherms of an equimolar mixture of $CO_2$ and $CH_4$ are reported in Fig. S4, and those of $CO_2$ and CO in Fig. S5. The $CO_2/CH_4$ selectivities at a total pressure of 70 bar are: 9.2 for $Br^-$, 10.2 for $Cl^-$ and 12.3 for $F^-$. This increase in selectivity with the hardness of the anion is independent of the forcefield used to describe $CO_2$, and could be reproduced with the original electrostatic TraPPE model (though the selectivities themselves are quite higher with TraPPE). This suggests that, of all three halogenide-Zn-CBTACN, F-Zn-CBTACN may be a particularly potent candidate for carbon dioxide/methane separation.

**Conclusions**

We have used high-pressure adsorption experiments and molecular simulation in a synergistic approach to study a novel family of cationic metal–organic frameworks, built from $Zn^{2+}$ cations, linked together by carboxylate-substituted triazamacrocycles. We developed a set of partial atomic charges for the description of the cationic framework, based on quantum chemistry calculations. These were used to predict the localization of extraframework halogenide anions within the nanopores, which is not available experimentally. High-pressure $CO_2$ adsorption experiments were performed, and their results used to optimize a $CO_2$ forcefield, based on the TraPPE model. We then predicted high pressure coadsorption properties of carbon dioxide/methane mixtures in members of the Zn-CBTACN family with various halogenide anions. We showed that the nature of anions influences separation properties, with the fluoride member of the family exhibiting a better $CO_2/CH_4$ selectivity than chloride and bromide variants.

**Acknowledgments.** We thank Guillaume Ortiz, Stéphane Brandes and Roger Guilard for a fruitful collaboration, discussions on their results and providing us with the Cl-Zn-CBTACN sample used in this work. This work was supported by the Agence Nationale de la Recherche Technologique and by the Agence Nationale de la Recherche under contract ANR-06-CO2-007. This work was performed using HPC resources from GENCI-IDRIS (grant 2010-086402).

**Supporting Information Available**. Structure files and snapshots, as well as additional pure-component adsorption and mixture coadsorption isotherms.

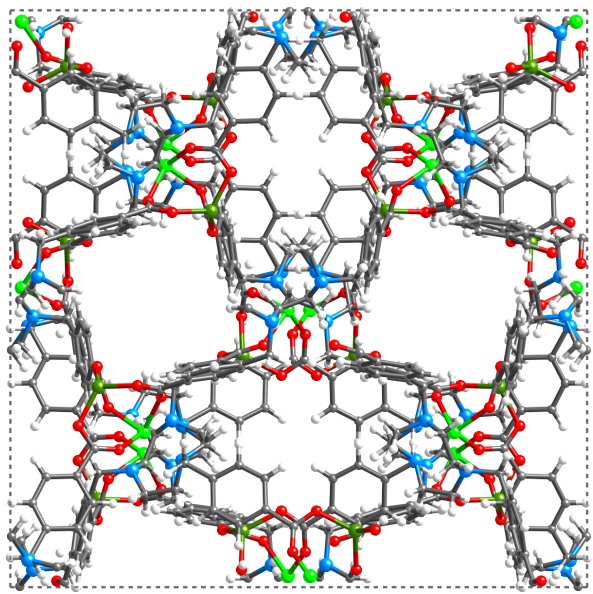

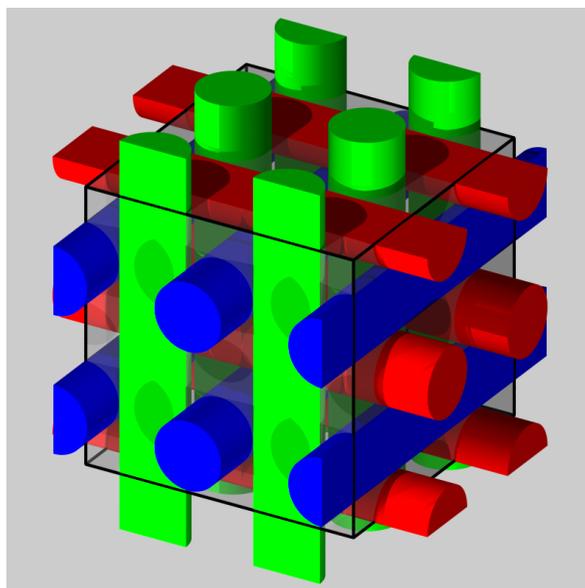

**Fig. 1:** Top: view of one unit cell of the Zn-CBTACN material (blue – N; green – Zn; red – O; gray – C; white – H). Bottom: schematic representation of the system of collinear and perpendicular channels of Zn-CBTACN in one unit cell.

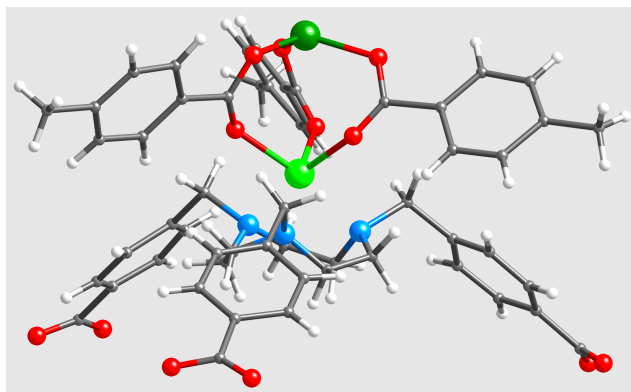

**Fig. 2:** View of the cluster of Zn-CBTACN from which the partial atomic charges for the material were calculated. See Figure S2 for an animated view of the cluster.

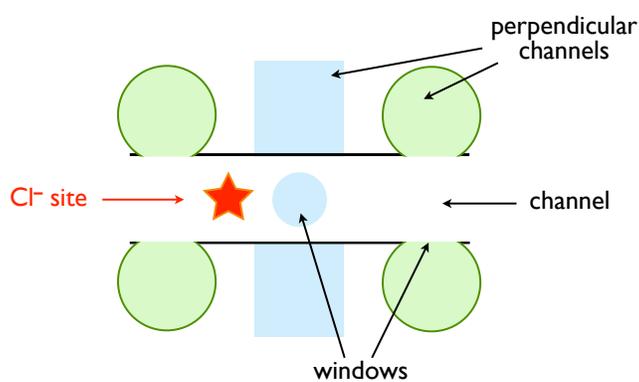

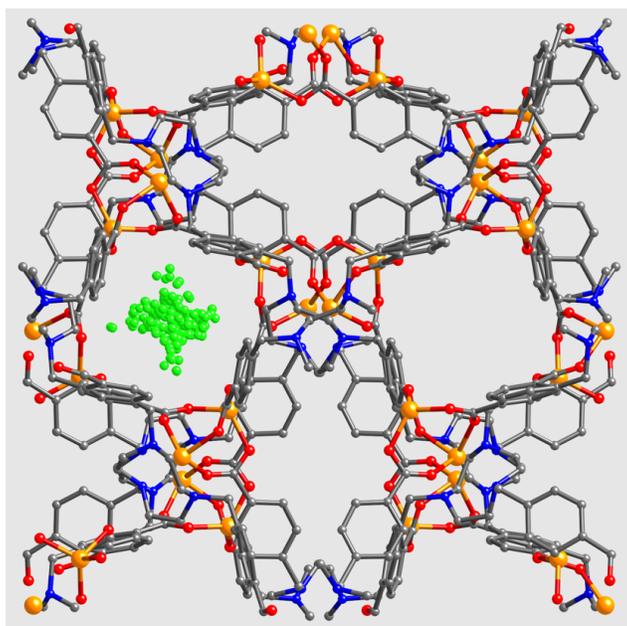

**Fig. 3:** Top: schematic view of the Cl⁻ site along one channel, in 2D. Bottom: picture of the Cl⁻ site along the channel (100) of the material.

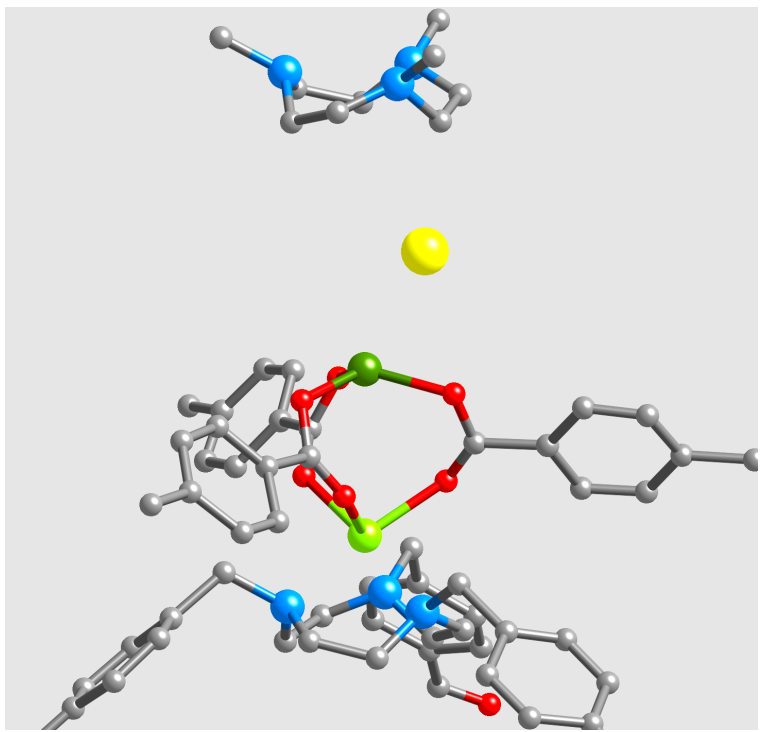

**Fig. 4:** Image of the F⁻ site (in yellow) next to an undercoordinated Zn (II) ion of the framework (dark green), and close to two nitrogen atoms of a neighbouring triazamacrocycle (on top).

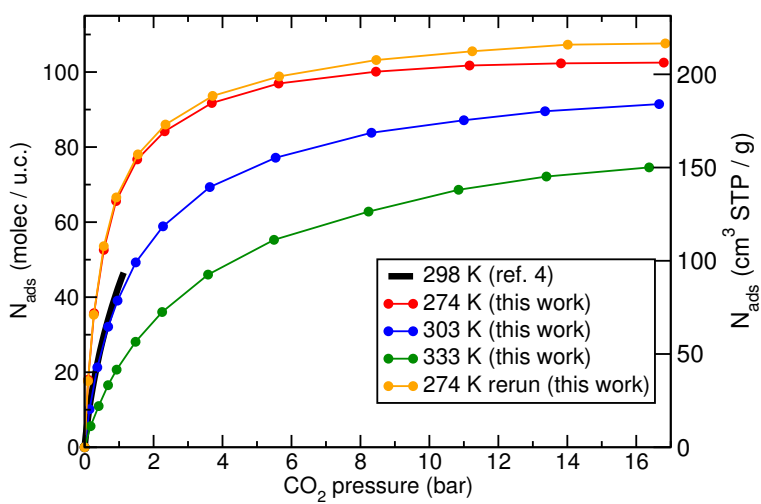

**Fig. 5:** Adsorption isotherms of $CO_2$ in Cl-Zn-CBTACN at 274 K (red), 303 K (blue) and 333 K (green) on the same sample. The orange curve represents a second isotherm taken at 274 K, measured after the 303 K and 333 K isotherms. In black, the adsorption isotherm from ref. 7 at 298 K.

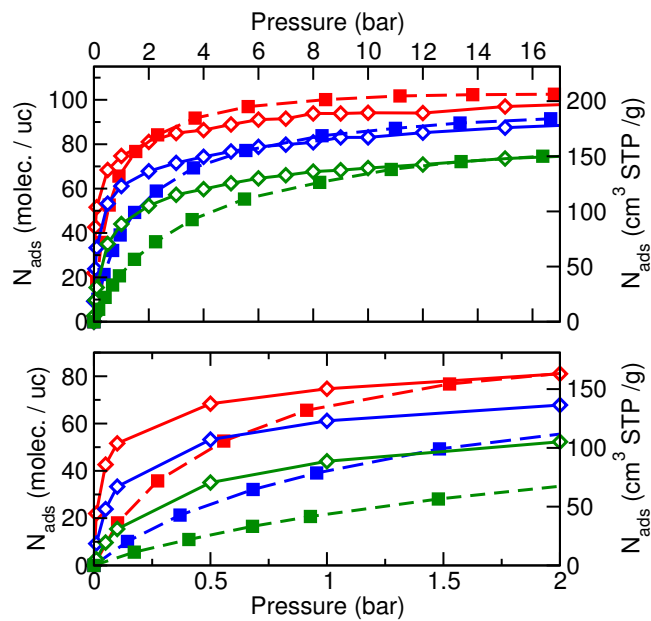

**Fig. 6:** $CO_2$ adsorption isotherms from molecular simulation using the TraPPE model of $CO_2$ (solid lines), compared to experiments (dashed lines) at 273 K (red), 300 K (blue) and 333 K (green).

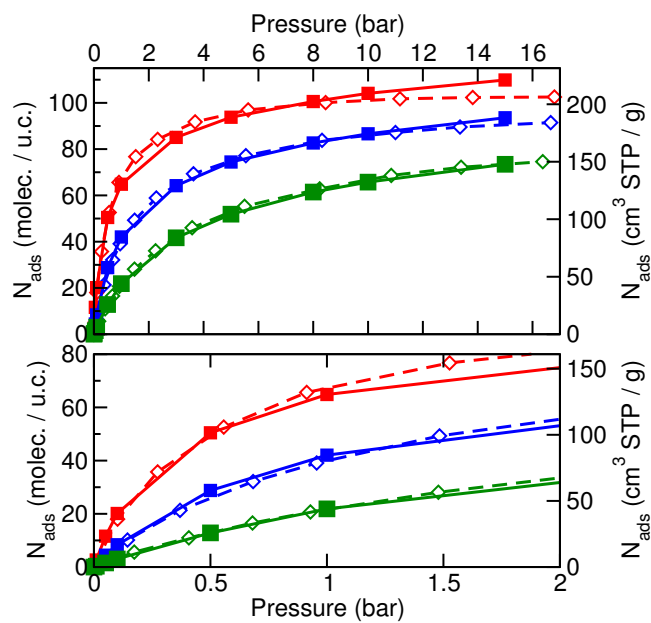

**Fig. 7:** Compared experimental (dashed line) and simulation (solid line) isotherms of $CO_2$ in Cl-Zn-CBTACN at 273 K (red), 303 K (blue) and 333 K (green), using the optimised nonelectrostatic forcefield.

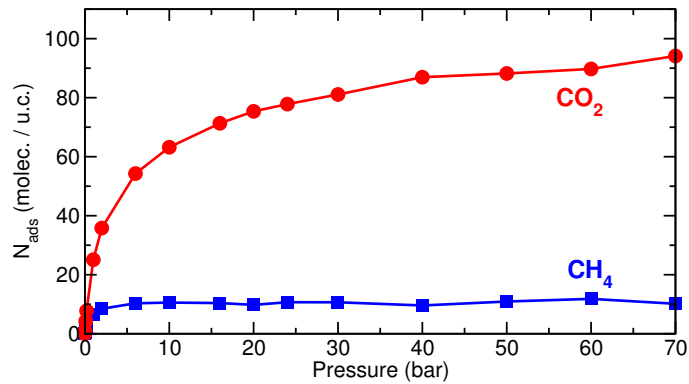

**Fig. 8:** Coadsorption isotherms of a 50:50 mixture of $CH_4$ and $CO_2$ in Cl-$Zn_2$-CBTACN family at 303 K.

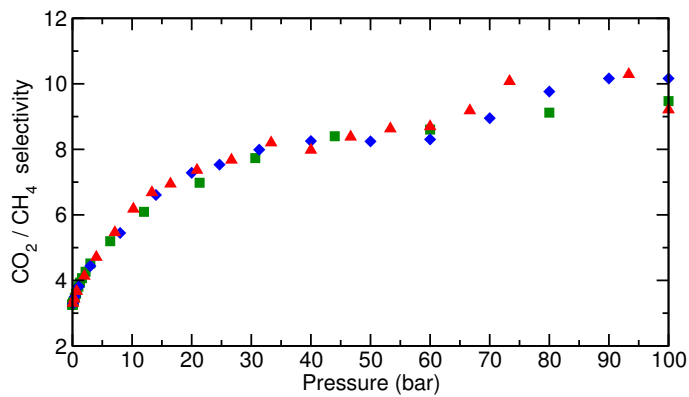

**Fig. 9:** Coadsorption selectivity of $CO_2/CH_4$ mixtures of various compositions in Cl-$Zn_2$-CBTACN at 303 K. Red: 25% $CH_4$; blue: 50% $CH_4$; green: 75% $CH_4$.

| Atom type | Localisation | Charge |
|---|---|---|
| N | Triazanonane | –0.02 $e$ |
| Zn (a) | Tetrahedral | +1.28 $e$ |
| Zn (b) | Octahedral | +0.75 $e$ |
| O | Carboxylate | –0.65 $e$ |
| C (a) | Carboxylate | +0.70 $e$ |
| C (b) | Triazanonane | –0.12 $e$ |
| H | Aromatic ring | +0.10 $e$ |
| All others | | 0 |

**Table 1:** Partial atomic charges for the metal–organic framework as determined by quantum chemistry calculations on a cluster.

| | TraPPE potential | | Adjusted potential | |
|---|---|---|---|---|
| | $\sigma_i$ | $\varepsilon_i$ | $\sigma_i$ | $\varepsilon_i$ |
| C | 2.80 Å | 27.0 K | 2.12 Å | 39.4 K |
| O | 3.05 Å | 79.0 K | 2.33 Å | 112.7 K |

**Table 2:** Lennard-Jones potential parameters used for $CO_2$ in this work: the standard TraPPE forcefield[12] and an readjusted nonelectrostatic potential accounting for better description of the Cl$^-$-$CO_2$ interactions.

**References.**

**Table of Contents Graphics**

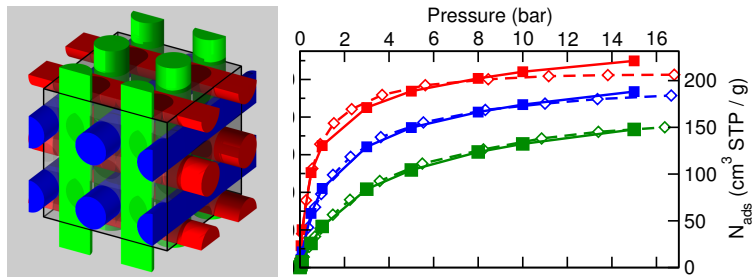